\documentclass[twocolumn,showpacs,showkeys,preprintnumbers]{revtex4}
\usepackage{graphicx}
\usepackage{dcolumn}
\usepackage{bm}
\usepackage{color}

\begin{document}

\title{Slope of the lateral density function of extensive air showers around the knee region as an indicator of shower age}
\author{Rajat K. Dey} 
\email{rkdey2007phy@rediffmail.com}
\author{Sandip Dam}
\email{sandip_dam@rediffmail.com}
\affiliation{Department of Physics, University of North Bengal, Siliguri, West Bengal, 734 013 India}

\begin{abstract}
Analyzing simulated extensive air shower (EAS) events generated with the Monte Carlo code CORSIKA, this paper critically studies the characteristics of lateral distribution of electrons in EAS around the knee energy region of the energy spectrum of primary cosmic rays. The study takes into account the issue of lateral shower age parameter as indicator of the stage of development of showers in the atmosphere. The correlation of lateral shower age parameter with other EAS observables is examined, using simulated data in the context of its possible use in a \emph{multi-parameter} study of EAS, with a view to obtaining information about the nature of the shower initiating primaries at sea level EAS experiments. It is shown that the observed slope of the lateral density function in the 3-dimensional plot at least for the KASCADE data supports the idea of a transition from light to heavy mass composition around the knee.
\end{abstract}

\maketitle

\section{Introduction}
The origin of high-energy cosmic rays (CRs) remains a mystery despite several attempts to unravel it over the last century. This situation is in part due to the inadequate knowledge of the nature and the spectra of energetic primary cosmic rays (PCRs) from their direct measurements above a few $10^{14}$ eV. Even the presently available sophisticated technology, viz. satellites and balloon borne experiments, cannot do the job because of the very low flux of the radiation. Currently the only feasible way to get information about such energetic particles is through the study of CR extensive air showers (EAS) which are cascades of many secondary particles produced when PCR particles interact with atmospheric nuclei during their advancement towards the ground.

There exists a power law behavior in the energy spectrum of PCRs with three breaks. First, a {\lq knee\rq} around $3$ PeV just after which the spectral index suddenly takes a value $-3.1$ from $-2.7$. Secondly, an {\lq ankle\rq} in the vicinity of the few EeV energy after which the spectrum flattens again to its original slope. Finally, at energies above $4\times 10^{19}$ eV, the PCR flux experiences a suppression [1-4] which is compatible with the Greisen-Zatsepin-Kuzmin (GZK) effect [5]. However, other possible reasons (e.g. upper limit in the energy at the source) for the suppression cannot be ruled out. Present understanding takes the knee to be a signature of galactic origin of PCRs. Hence, measurements of several characteristics of PCRs including its mass composition around the knee are very important for proper understanding of the origin of this spectral feature. 

Observables from EAS experiments are known to be densities and arrival times of cascade particles. More than $90\%$ of the various charged secondaries are electrons (sum of $e^{-}$ and $e^{+}$ henceforth termed as electrons). These observables are measured at different radial distances, and they contain all information on the primary particle. By applying the shower reconstruction to the electron density data using the so called Nishimura-Kamata-Greisen (NKG) type lateral density function (LDF), the crucial EAS parameters such as shower size ($N_{\rm e}$), the EAS core and the shower age ($s$) are obtained. In EAS experiments equipped with {\lq hybrid\rq} detector system [6-7], EAS components such as muons ($\mu^{\pm}$), hadrons ($h\bar{h}$), air shower associated Cherenkov photons, and fluorescence light are also measured simultaneously with electrons. The observed results are usually interpreted by means of detailed Monte Carlo simulations which cautiously account various interaction processes of primary or a few secondary particles as input. But our understanding of hadronic interactions at high energies is limited. In addition, there exists huge fluctuations in interaction mechanisms in a cascade which however are intrinsic to any EAS. Therefore, any conclusion derived from EAS measurements remains somewhat uncertain, and even sometimes there are huge inconsistencies on conclusions drawn from the experiments. Therefore, a multi-parameter approach in this context is expected to be more efficient. In this approach, several possible EAS parameters are simultaneously considered in order to extract information on shower initiating primaries in EAS experiments with smaller uncertainties.

In the present work, the shower age $s$, which is directly linked to the slope of the LDF (NKG), and essentially describes the form or shape of the lateral distribution of EAS electrons, is adjudged an important parameter in the multi-parameter approach. It provides the fundamental relation between the depth of shower maximum ($X_{\rm max}$) and the slope of the LDF. It also determines the stage of shower development, starting mainly from the $X_{\rm max}$ to the ground as per electromagnetic (EM) cascade theory. Hence, the value of $s$ is related to the degree of steepness of the lateral distribution of electrons. By virtue of these features of the $s$ parameter, it is still being used to study the PCRs.

We organize the paper as follows. In the next section, we discuss how the concept of shower age, introduced in the EM theory, is used to describe the evolution of an EAS in the longitudinal and transverse (i.e. lateral) directions with respect to the shower axis in the atmosphere. In section III, the characteristics of the method of simulation used in this work will be described. The same section also includes the method of analysis of simulated data for estimation of air shower parameters. The results of the study are presented in section IV with the necessary discussion. Conclusions will be drawn in section V. 

\section{The concept of shower age describing the extensive air shower development}												
The concept of the shower age has been used in CR studies since around 1940. It first appears in the early works of Rossi and Greisen [8], and later in Nishimura's work [9]. Their works exploited the concept in formulating some expressions needed for the solutions of the diffusion equations due to longitudinal development in the EM cascade theory. It was then applied to the lateral spread of electrons around the cascade axis while solving the diffusion equations in the three-dimensional EM cascade theory. 

The most simple and earliest description of the EM cascade was being started with the longitudinal development. One of the most popular analytic representations of the average longitudinal profile was provided by Greisen [10]. This so called Greisen profile was obtained from the calculations carried out by Snyder in the so called approximation B (taking into account only three processes, (a) pair production by the photons, (b) bremsstrahlung by the electrons and, (c) collision energy loss suffered by the electrons), is given by      
\begin{equation}
~{N_{\rm e}}~ =~ \frac{0.31}{\sqrt{\ln(E_{0}/\epsilon_{0})}}~\exp[X(1-1.5~\ln(s_{\parallel})], 
\end{equation}
where $s_{\parallel}$ gives the stage of the longitudinal cascade development, $\rm E_{0}$ is the energy of the primary photon generating the cascade (the atmospheric depth $X$ is expressed in cascade units; it is normalized by the radiation length of electrons in air, that is equal to 37.1 g cm$^{-2}$).  

The average evolving stage of pure EM cascades is expressed primarily by the longitudinal shape or shower age $s_{\parallel}$ parameter. It is essentially the inverse of the fractional rate of change of the total electron number of an EAS with atmospheric depth [8]. This fractional rate of change is called the size slope $\lambda(s_{\parallel})$ as    
\begin{equation}
\lambda (s_{\parallel}) =\frac{1}{N_{\rm e}(X)}\frac {{\rm d}N_{\rm e}(X)}{{\rm d}X}.   
\end{equation}
Thus, the shower age is simply related to the rate of growth and decay of $N_{\rm e}(X)$ in a shower, and mathematically expressed by the operation ${\lambda}^{-1}(\frac{1}{N_{\rm e}(X)}\frac {{\rm d}N_{\rm e}(X)}{{\rm d}X})$.

After doing a few mathematical operations to the equation (2), Rossi and Greisen attributed a more direct physical meaning to the shower age as a shape or spectral index of the energy spectra of EM particles. In a shower initiated by an electron or photon, the energy spectra of secondary electrons or photons follow a power law of the form,  
\begin{equation}
{n_{\rm e/{\gamma}}(E)} \sim E^{-(s+1)}.
\end{equation}
The power law works above the critical energy $\epsilon_0$, where $\epsilon_0$ is the energy value at which ionization and bremsstrahlung energy losses are equal, by an electron, which happens at $\epsilon_0 \approx 82$ MeV for electrons in air.  

After the preliminary work of Moli\'ere and Bethe, focusing on the lateral electron distribution near the cascade maximum, Nishimura and Kamata described the lateral development under the assumption that electrons suffer constant amount (energy independent) of collision loss  per radiation length, which is equal to the critical energy (approximation B) [11] by solving the 3D diffusion equations. Fitting their numerical results on the electron densities, Nishimura and Kamata were able to propose a first $r$-dependence analytic expression for the shower age $s_{\parallel}$
\begin{equation}
s_{\parallel}=\frac {3X}{X+2\ln(E/\epsilon_{0})+ 2\ln(r/r_{\rm m})},   
\end{equation}
where $X$ is the atmospheric slant depth (in electron radiation length unit) of the observational level, $E$ is the energy of the primary electron/gamma; $r$ and $r_{\rm m}$ are the radial distance from the shower core, and the Moli\'ere radius ($r_{\rm m}$ corresponds the radius of a cylindrical region where $\sim 90\%$ of the EM shower energy is deposited, that is, $r_{\rm m}\sim 80$ m at sea level). With this definition, $s_{\parallel} = 0$ at the top of the atmosphere and $s_{\parallel} \sim 1$ at the depth of shower maximum. Under the same theoretical framework, the LDF of cascade particles advancing through a constant air medium, was expressed approximately by Greisen, and is very well-known as the Nishimura-Kamata-Greisen (NKG) structure function [10], given by
\begin{equation}
{\rm f}(r)={\rm C}(s_{\perp})(r/r_{\rm m})^{s_{\perp}-2}(1+r/r_{\rm m})^{s_{\perp}-4.5} \;,
\end{equation}
where $s_{\perp}$ essentially describes the slope of the LDF which is determined from the shape of the observed lateral distribution of electrons in an EAS. The normalization factor C($s_{\perp}$) in the equation (5) is given by 
\begin{equation}
{\rm C}(s_{\perp}) = \frac{\Gamma(4.5-s_{\perp})}{2\pi\Gamma(s_{\perp}) \Gamma(4.5-2s_{\perp})}  \;.
\end{equation}

An expression for the electron density ($\rho_{\rm e}$) in terms of the profile function ${\rm f}(r)$ (thanks to the properties of the Eulerian function as shown in appendix: section 3) is,
\begin{equation}
\rho_{\rm e}(r) = {N_{\rm e} \over r_{\rm m}^2 }{\rm f}(r) \;.
\end{equation} 

The relation $s_{\parallel} = s_{\perp}$ holds for EM showers [11]. Such equivalence implies the correlation between steeper lateral distributions ($s_{\parallel} < 1$) or flatter lateral distributions ($s_{\parallel} > 1$) in proportion to the distance to shower maximum. It suggests also the employment of the parameter $s_{\perp}$ as a hint of the global shower cascading expected to depend on shower initiating particle and several interaction features (cross-section, multiplicity and in-elasticity).

However, Nishimura, Kamata and Greisen noted some difficulties when comparing the prediction of the cascade theory with the measured lateral  distributions of electrons in EAS. We have listed hereunder some limitations on the validity of the analytical expressions caused by the various approximations (so-called approximations A and B) in obtaining the solutions as well as due to over-simplification of the adopted 3D transport equations. The validity of the equation (5) therefore obeys some conditions, which are as follows:
\begin{itemize}
\item [-] $E_{0}/ \epsilon_{0} \gg 1$; due to the asymptotic value of the cross sections
\item [-] $X > 1$; more than one radiation length unit is necessary to consider the continuity in the cascade
\item [-] $\mid\log(E_{0}/\epsilon_{0})\mid \gg \mid\log(r/r_{\rm m})\mid$
\end{itemize}  
 
The last condition may be not satisfied in the vicinity of the EAS axis as underlined by Nishimura considering not appropriate to use the formula near the EAS core. Nishimura and Kamata also ascertained that corrections were necessary to their calculations performed for the homogeneous medium. In consequence, taking into account the variation of the air density, they inferred that in order to compare the theoretical curves with the experimental data, the distribution obtained at about 2 radiation length (with approximation B) above the measurement depth must be used. 

Solutions of the 3D diffusion equations came from simultaneous operations such as solutions of adjoining equations, the Bessel Fourier transformation being applied to $r$, the numerical integration on $X$, and on the energy $E$ being performed by partial polynomial functions [12-13]. This method, which involves several approximate steps and numerical inverse transformations, leads undoubtedly to a limiting (as $E_{0} \rightarrow \infty$) lateral distribution of EAS electrons, e.g. the NKG LDF. The meaning of \emph{limiting} actually refers to the energy condition i.e., $E_{0}$ or $E >> \epsilon_0$ through the saddle point solutions in approximation A, and the negligible collision losses of electrons in approximation B, that have been used to obtain the NKG LDF.

The equation (1) concerns the total number of electrons of the cascade, whereas a parallel relation for the integral energy spectrum of the electrons is also given; in both cases cross sections for bremsstrahlung and pair production in their respective asymptotic form at very high energy have been exploited. In this useful synthesis the $s_{\parallel}$ parameter is defined simply as 
\begin{equation}
s_{\parallel} = \frac{3 X} { X + 2 X_{\rm max}}.
\end{equation}

So far the concept of the shower age is elaborated explicitly from the angle of purely EM cascade theory. An extension of the concept is applied also to hadron initiated showers. It was Cocconi who first pointed out that the equations (1), (5) and (8) provide a complete procedure for calculating the lateral development of a pure EM cascade [14]. A hadronic shower is not a pure EM cascade but consists of a superposition of many pure independent EM cascades that usually build the lateral distribution of electrons of the shower. Hence, a resulting single cascade with a modified shape parameter will effectively describe both the longitudinal and lateral structures of electrons for the hadron initiated shower [11]. This is a possible reason why the analytical expressions in (1), (5) and (8) from the EM theory are only rough approximations for the description of hadron initiated showers.

In some earlier analyses of observed EAS data, it was found that electron densities at large core distances were noted to be larger than NKG predicted values [6-7]. The muon decay process might be a possible reason for the origin of it. On the other hand, the original NK (Nishimura-Kamata) predicted electron densities near the core were found a little higher than those with the NKG in the domain $\rm s \geq 1.2$ [15]. It was also noted in Monte Carlo simulations that the NKG formula was quite useful for proton initiated showers and is much faster than the full simulation of the EGS cascade [16]. 

In EAS studies a significant part of various discrepancies among theory, simulation and experiment is of course due to the application of a pure EM cascade (NKG type LDF) to hadronic showers. But, discrepancies might also come either from the simplistic treatment of scattering processes (e.g. the multiple coulomb scattering and also Bhaba, and Moller scattering) in NKG or neglecting of effects like photo-production, geomagnetic deflection. Also hadronic interactions affect the shape of the lateral distribution of electrons to some extend. 

The limitations of various instruments used in an EAS experiment may also sometimes give rise to discrepancy. For example, while detecting large showers in KASCADE experiment (suppose, the setup is appropriate to observe medium size showers), the detectors close to the EAS core get saturated and should not be considered for the analysis. Then, the LDF does not provide a better fit for large showers at smaller distances and for small showers at larger distances [17].      

Many authors therefore replace the NKG type LDF with slightly different functions [18], or modify the LDF by changing the parameters of the exponents to account for the hadronic origin of showers in EAS data. The second treatment was indeed implemented by the KASCADE-Grande collaboration to provide a better agreement to the shape of the observed lateral distribution of electrons with the corrected NKG function [17].

In Tien Shah experiment, a steeper profile was exhibited near the core by electron density data, and it was suggested that the NKG LDF be corrected by replacing the Moli\'ere radius $\rm r_{0}$ with $0.6\rm  r_{0}$ for a better agreement [19]. 

The Monte Carlo calculations of Hillas [20] were used to improve the NKG LDF by simultaneous modifications of the exponents and the Moli\'ere unit to achieve better agreement with data. It was suggested by some authors that variation in the Moli\'ere unit, instead of the exponents, will lead to a better improvement for the fit behaviour of the NKG function. This was subsequently implemented to the NKG function by Uchaikin and later the KASCADE group by replacing the Moli\'ere radius with a longitudinal age dependent effective Moli\'ere radius [12-13,21]. The Uchaikin LDF has restricted applicability, it does not work well to fit individual shower owing to large fluctuations between the detector locations. This modified NKG LDF was actually implemented in the so called NKG subroutine of CORSIKA for the analytical treatment of EM component in a shower. 

A theoretical prediction was made by Capdevielle with regard to the errors introduced by the NKG function in shape and shower size determination [22]. It was then supported by many observations (e.g. in [23]), and subsequently concluded by Capdevielle et al. [22,24] that the NKG function with a single age cannot describe the lateral distribution of electrons. Consequently, the notion of \emph{local age} emerged [16,22].

Attempts were made by several EAS experiments [12-13,17-20] to assign important physical significance to the parameter $s_{\perp}$, that is obtained from the fit on the observed density data of electrons, as predicted by the EM cascade theory. Some experiments such as GRAPES-3 [25], ARGO-YBJ [26] etc. have realized that it could be a potential observable to study the PCRs. In the cascade theory, the parameter is supposed to relate with the stage of development of EAS in the atmosphere [11]. Distribution of secondary electrons either in spatial or energy space on the ground depends only on the stage of longitudinal shower development in the atmosphere or equivalently on $s_{\parallel}$. An equality between these two shower age parameters exists in the cascade theory for pure EM showers. However, such equality is also maintained approximately by hadron induced showers [11,27].

Recent studies indicate that the average shape of a number of distributions of electrons in showers initiated by hadrons exhibit the so called \emph{universality} [28-33]. The fact that the \emph{universality} property of air showers [34-35] is such a powerful tool for the analysis of EAS data can also be understood from the essence of the concept of shower age. According to the \emph{universality} property, all showers follow a similar development in the vicinity of their shower maxima, where they also share the same shower age. These showers, for an equal density profile of the medium also have the same lateral spread around the cascade axis. 

It was also demonstrated [11,27] that for hadron initiated EAS, the longitudinal and lateral profiles of the EM component can be described by some average overall single cascade, assigning a suitable value to the age parameter. The so-called universality property of simulated showers has been established for EAS induced by primaries at ultra-high energies in terms of the shape parameter $s_{\parallel}$ [28-29,31] using $X_{\rm max}$. However, observed lateral density distribution of electrons is usually described in terms of $s_{\perp}$ and in most experiments the estimated lateral/transverse shower age i.e. the $s_{\perp}$ differs from the $s_{\parallel}$ for EAS with hadrons as primaries. It was suggested from the experimental results [19,36] that these two age parameters are connected through the (approximate) relation $s_{\parallel}\geq s_{\perp} + \delta$, with $\delta \approx 0.2$. Some early Monte Carlo simulation results obtained a relation of the nature $s_{\parallel}\sim 1.3s_{\perp}$ [37]. The $s_{\parallel}$ parameter can be estimated observationally only if EAS experiments are equipped with Cherenkov or fluorescence detectors, whereas the $s_{\perp}$ parameter follows straightway from the LDF for electrons which is the basic measurement of any conventional EAS array consisting of particle/scintillation detectors.

In this paper, a correlation between $s_{\perp}$ and the stage of the longitudinal development will be examined using simulated EAS data according to the prediction from the EM cascade theory. Next, we investigate various characteristic features of the parameter and its relationship with some other important EAS parameters using Monte Carlo simulations. In this context, we have to consider some observed results from two sea level experiments i.e. NBU and KASCADE for the purpose of comparison and interpretation. The NBU air shower array is located at North Bengal University campus, India (latitude $26^{\rm o} 42'$ N, longitude $88^{\rm o} 21'$ E, 150 m a.s.l., area 2000 m$^2$), was being operated during 1980 - 98 [38]. The KASCADE experiment is located at Forschungszentrum Karlsruhe, Germany (latitude $49.1^{\rm o}$ N, longitude $8.4^{\rm o}$ E, 110 m a.s.l., area 40000 m$^2$) [39]. Our main objective is to explore the significant contribution of $s_{\perp}$ in a  so-called multi-parameter approach of studying EAS data to derive possible conclusions on the nature of shower initiating particles around the knee region. 

We have used the original form of the NKG LDF to fit the density of electrons whereas in the KASCADE work the modified form of the NKG was used [17]. Moreover, in their work, two secondary estimators were used for the measurement of energy and mass. The nature of variation of the shape parameter was studied as a function of these estimators independently. In the present work, we show that the slope of the LDF of EAS electron distribution from a detailed Monte Carlo simulation study indicates the stage of development of showers in terms of $s_{\perp}$. We also investigate the fluctuations in $s_{\perp}$ arising from its frequency distributions and  the correlation of the fluctuation parameter with $N_{\rm e}$ for proton and iron showers. An important correlation of $s_{\perp}$ simultaneously with $N_{\rm e}$ and $N_{\mu}^{\rm tr.}$ ($N_{\mu}^{\rm tr.}$ takes into account the total number of muons from a defined radial bin, e.g. $4\textendash 35$ m at NBU level) are studied in a so called multi-parameter approach. Efforts are also made to explore whether the $s_{\perp}$ has any correlation with the $s_{\parallel}$ parameter. It is expected that a more direct application of these parameters from the original LDF or Greisen profile function to simulated or observed data will provide an opportunity to re-examine the important predictions in the EM cascade theory better. For generating Monte Carlo shower events, the air shower simulation program CORSIKA (COsmic Ray SImulation for KAscade) version 6.970 [40] has been used in the work.     

\section{Simulation and data analysis}
The hadronic interaction model QGSJet01.c [41] is considered for interaction processes above $80 {\rm GeV/n}$ energy. The model is combined with the low energy hadronic model GHEISHA version 2002d [42] that works below $80 {\rm GeV/n}$. All these models are embedded in the CORSIKA Monte Carlo program version 6.970 [40]. Although the low energy interaction model GHEISHA has some shortcomings but it does not have significant effect on the secondary particle distribution except at very large core distances [43]. For the EM part the EGS4 [44] program package has been embedded in the Monte Carlo code for carrying out interactions in the EM component. The high-energy hadronic model SIBYLL version 2.1 [45] is considered to check the potential impact, of the high-energy hadronic models on some of our important findings. 

In the CORSIKA Monte Carlo code, the US-standard atmospheric model [46-47] with planar approximation is used for the generation of EAS events with zenith angles $\Theta < 70^{\rm o}$. We have simulated events with $\Theta$ that ranges between $0^{\rm o}$ and $50^{\rm o}$. The EAS events have been generated for proton, helium and iron nuclei as primaries in the primary energy interval of $10^{14}$ eV to $3 \times 10^{16}$ eV with about $20000$ events for each primary. The spectral index of the primary energy spectrum has been taken as $-2.7$ up to the knee ($3 \times 10^{15}$ eV) and as $-3.1$ beyond the knee in the simulation. A mixed composition sample has also been prepared from the generated showers containing $50 \%$ proton, $25 \%$ helium and $25 \%$ iron showers. The simulated showers have been generated at the altitude and geomagnetic fields corresponding to the North Bengal University air shower array. At the ground plane the kinetic energy cut-offs are set at 50 MeV and 3 MeV for muons and electrons. A smaller sample of simulated events has also been generated for KASCADE conditions to observe the nature of variation of the $s_{\perp}$ with other observables. The kinetic energy cut-offs are set at 230 MeV and 3 MeV for muons and electrons for KASCADE [17]. 

The simulated densities for electrons at different radial distances from the EAS core are used for shower fit, employing the NKG type LDF for obtaining $s_{\perp}$ parameter. It should be however mentioned that other EAS parameters such as $N_{\rm e}$, $N_{\mu}$, $\Theta$ and $\Phi$ are used directly from the generated events. Only the $s_{\perp}$ parameter is obtained from the fit using simulated/observed data. The KASCADE data for $\rho_{\rm e}$ are obtained from published papers corresponding to different $N_{\rm e}$ bins [21].  

In order to check the effectiveness of the NKG LDF in describing the simulated density of electrons, we have applied the fit procedure to the simulated density data. The estimation of the $s_{\perp}$ parameter is performed by means of a chi-square minimization routine using gradient search technique. In this procedure the simulated electron densities in the radial interval $7-102$ m are compared with the NKG LDF. It is found that the NKG function represents the simulated data satisfactorily over the whole radial distance except at very small distances, where the simulated densities are found a little higher than that given by the NKG function. Hence, the simulated densities only in the radial interval $7-102$ m are used, and fitted showers with reduced $\chi^{2}$ less than $4$ are only accepted for results. For primary energies around $10^{14}$ eV, the extension of a shower is usually spread over within $\sim 40$ m core distance whereas it exceeds $\sim 100$ m for energies around the knee. In the Figure 1, we have plotted the simulated electron densities over the range $7-102$ m for a particular $N_{\rm e}$ obtained with the QGSJet model and its comparison with the NKG fitted curve.

\begin{figure}
\centering
\includegraphics[width=0.5\textwidth,clip]{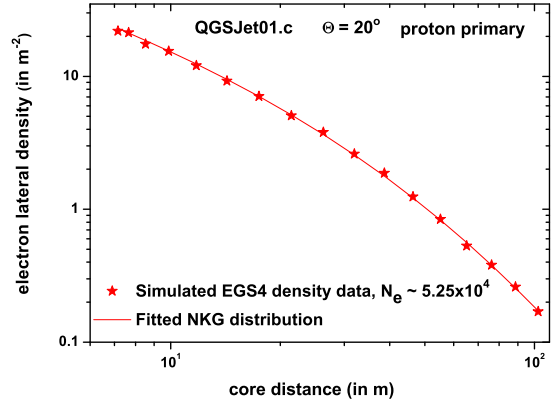} \hfill
\caption{Comparison of the NKG fitted lateral distribution of electrons with the simulated data for showers generated at $5\times 10^{14}$ eV energy.}
\end{figure}

The errors of the $s_{\perp}$ estimation in the interval, $N_{\rm e}: 10^{3}{\textendash}10^{5}$ are obtained as $\pm 0.034$ for the QGSJet01.c and $\pm 0.052$ for SIBYLL v.2.1. The SIBYLL model shows relatively higher uncertainty of the $s_{\perp}$ estimation and that might have originated from comparatively a smaller sample size of simulated data in this case.

\section{Results and discussions}
\subsection{Dependence of lateral shower age on atmospheric depth}
With the increase of $\Theta$, a shower traverses an increased thickness of atmosphere which immediately suggests that the EAS with higher $\Theta$ should be older in shower age than the EAS of smaller $\Theta$ but of same primary energy. Based on this idea, we have studied variation of $s_{\perp}$ with atmospheric depth from the simulated data. 

Figures 2a, 2b and 2c show the $s_{\perp}$ as a function of atmospheric depth for three different intervals of shower size, $(1.8\textendash 3) \times 10^{3}$, $(5.7\textendash 9.5) \times 10^{3}$ and $(2.0\textendash 5.0) \times 10^{4}$, obtained from Monte Carlo simulations. Separate samples of proton and iron initiated showers are considered to investigate the mass dependence. Besides, we have used our prepared mixed composition sample as described in the section III for this study as well. In the Figure 2c, we have included the NBU data [48].\\

\begin{figure}
\centering
\includegraphics[width=0.5\textwidth,clip]{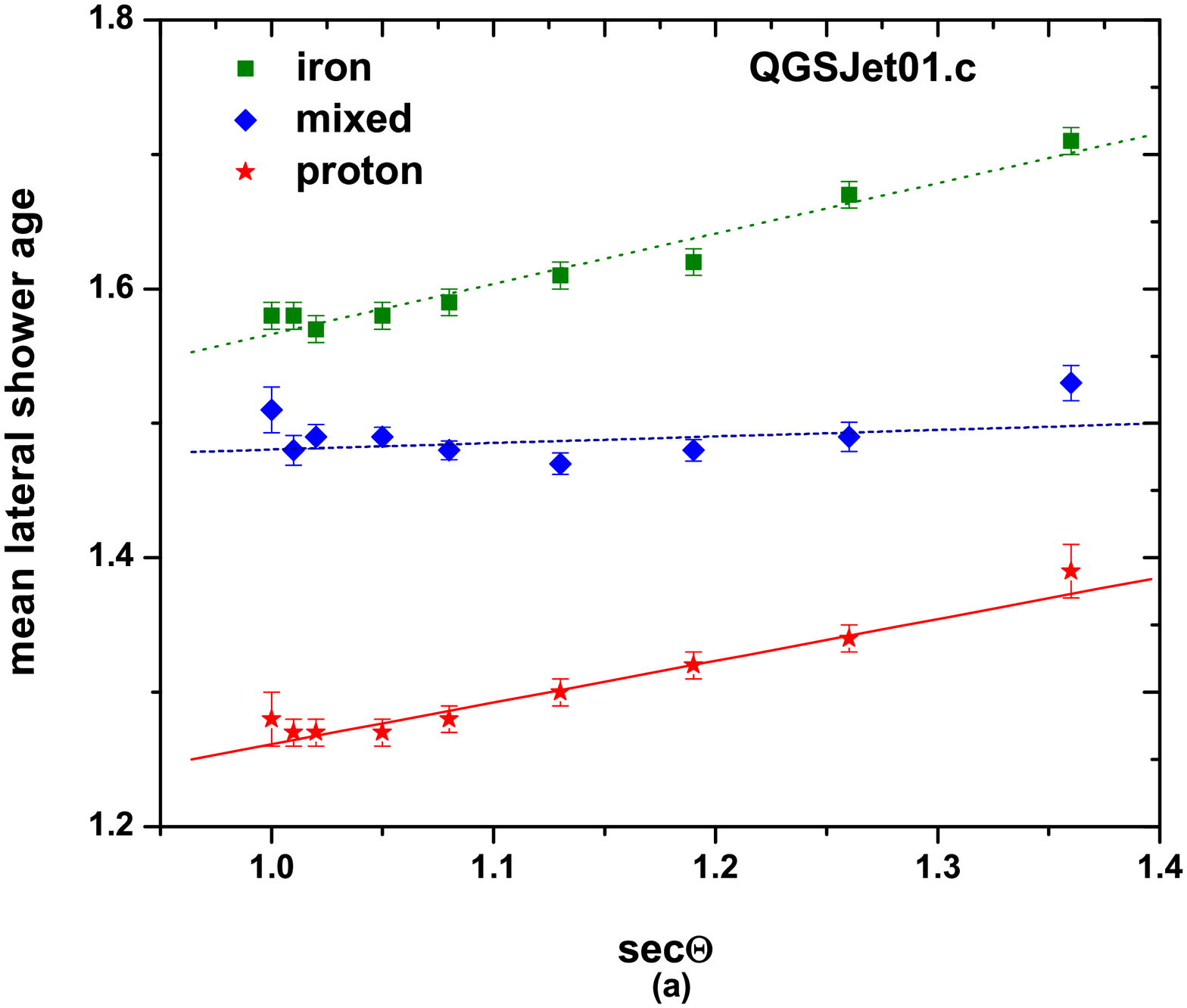} \hfill 
\includegraphics[width=0.5\textwidth,clip]{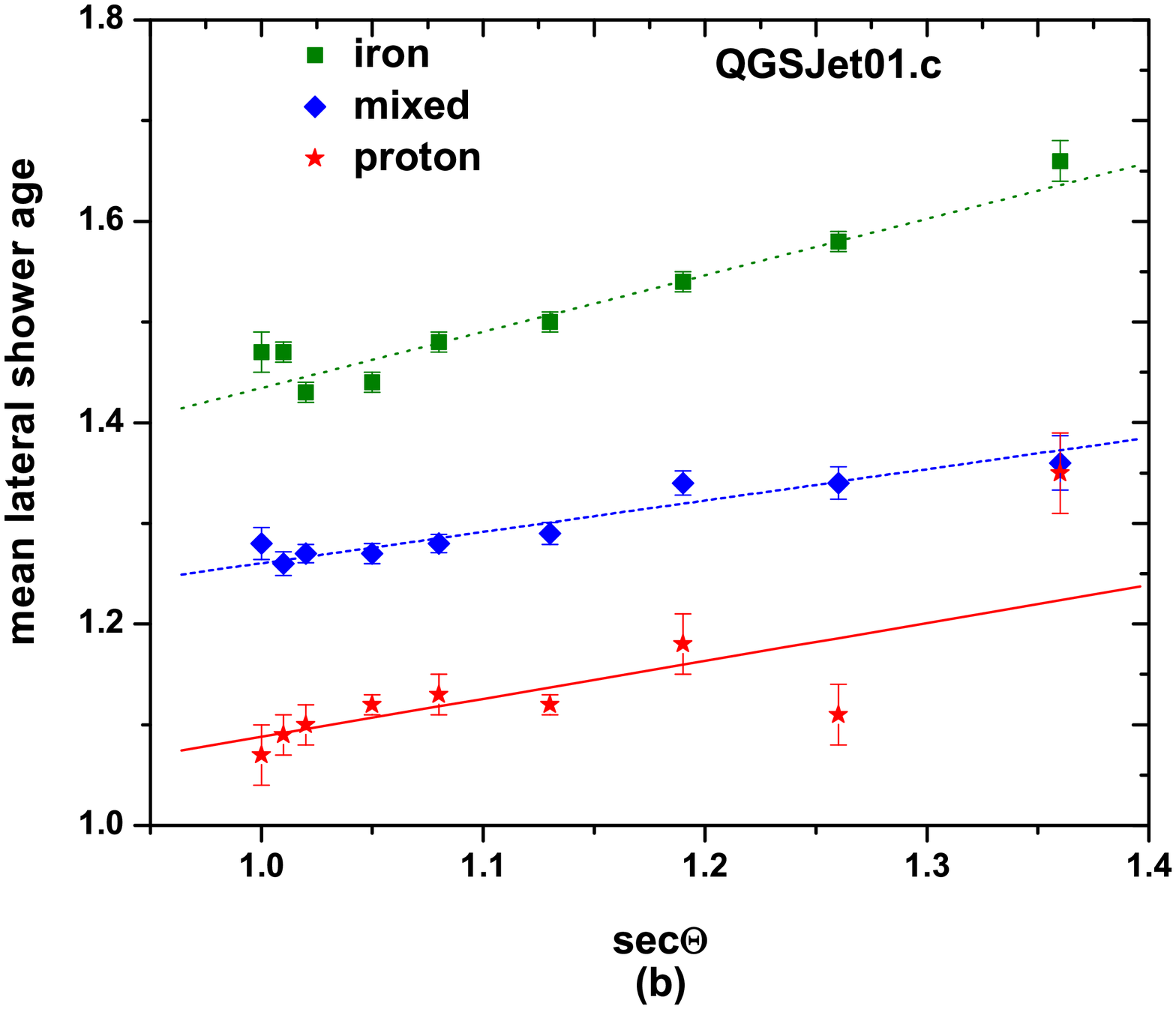} \hfill
\includegraphics[width=0.5\textwidth,clip]{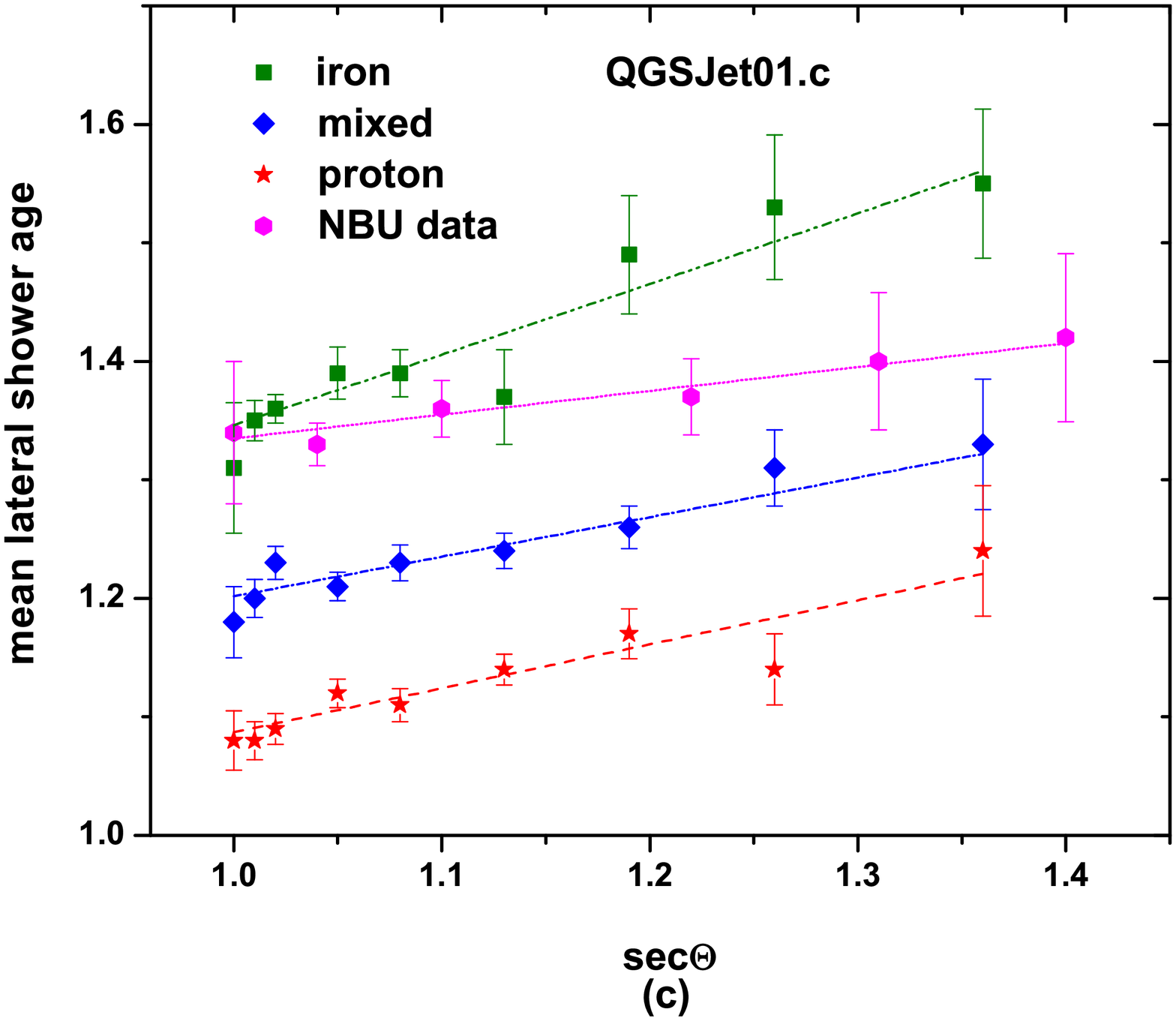} \hfill
\caption{Variation of shower age parameter ($s_{\perp}$) with atmospheric slant depth ($\sec\Theta$) for three $N_{\rm e}$ ranges: (a) $(1.8\textendash 3.0) \times 10^3$; (b) $(5.7\textendash 9.5) \times 10^3$ and (c) $(2.0\textendash 5.0) \times 10^4$ at NBU level. In (c) NBU data are provided.}
\end{figure}

\begin{table*}
\begin{center}
\begin{tabular}
{|l|l|l|l|r|} \hline

\multicolumn{2}{|c|} {$N_{\rm e} \times 10^3$} & $1.8-3.0$  & $5.7-9.5$ & $20-50$   

\\ \hline

proton  &\multicolumn{1}{c|}{\em$s_{\rm o}$}   & $0.95 \pm .03$   &$0.71 \pm .15$  & $0.71 \pm .09$  \\
      
   &\multicolumn{1}{c|}{\em$\rm A$}       & $0.31 \pm .02$   & $0.37 \pm .14$ & $0.37 \pm .08$  \\ \hline      

iron &\multicolumn{1}{c|}{\em$s_{\rm o}$}   & $1.19 \pm .03$   & $0.87 \pm .07$ & $0.74 \pm .13$  \\
      
   & \multicolumn{1}{c|}{\em$\rm A$}      & $0.37 \pm .03$   & $0.50 \pm .08$  & $0.59 \pm .12$ \\ \hline      

mixed  &\multicolumn{1}{c|}{\em $s_{\rm o}$} & $1.43 \pm .05$ & $0.95 \pm .05$ &$0.87\pm .08$ \\
      
       &\multicolumn{1}{c|}{\em $\rm A$}     & $0.04 \pm .04$ & $0.31 \pm .04$ & $0.33\pm .07$ \\ \hline 
NBU data  &\multicolumn{1}{c|}{\em $s_{\rm o}$} & $\textendash$ & $\textendash$ &$1.13 \pm 0.04$ \\
      
       &\multicolumn{1}{c|}{\em $\rm A$}     & $\textendash$ & $\textendash$ & $0.20 \pm 0.03$ \\ \hline 			
					
\end{tabular}
\caption {Parameters from the variation of $s_{\perp}$ with $\sec\Theta$ for three different $N_{\rm e}$ intervals. The NBU data are shown for one $N_{\rm e}$ interval only.} 
\end{center}
\end{table*}

It can be seen from the figures that for both proton and iron initiated showers, $s_{\perp}$ monotonically increases with atmospheric depth reflecting a strong correlation of the parameter with longitudinal development. The nature of $s_{\perp}$ versus atmospheric depth ($\sec\Theta$) variation does not have any noticeable dependence on shower size in the considered shower size range. 

For mixed primaries the $s_{\perp}$ parameter is found to vary relatively slowly with atmospheric depth and even noticed to remain nearly constant at some $N_{\rm e}$ intervals (see Figure 2a). This is due to the fact that showers initiated by heavier primaries dominate in the mixed composition sample when $\sec\Theta$ values are small. But, showers initiated by lighter primaries contribute most in the mixed sample for larger values of $\Theta$ in the same $N_{\rm e}$ interval. Similar explanation is valid also to the observed variation in the Figure 2c. However, NBU data show relatively higher values for $s_{\perp}$ in comparison with mixed data. Such a behavior from observation is obvious, because the observed shower events may contain species other than $\rm Z = 1, 2, 26$. Contamination of species with $Z > 2$ in NBU data might possibly push the $s_{\perp}$ versus $\sec\Theta$ curve close to the simulated iron curve.
 
The present simulation study suggests that the $s_{\perp}$ parameter can be expressed as a function of $\Theta$ by the empirical relation
\begin{equation}   
s_{\perp} = s_{\rm o} + {\rm A} \; sec \;\Theta.
\end{equation}
The values of $s_{0}$ and $\rm A$ for different $N_{\rm e}$ ranges obtained by fitting the simulated data generated with the QGSJet model are shown in Table 1. 

Equation (9) can be written as $s_{\perp} = s_{\rm o} + {\rm A} \; \frac{X}{X_{\rm v}}$, where $X$ and $X_{\rm v}$ are the atmospheric thickness traversed by the EAS and the vertical atmospheric depth respectively. It thus immediately follows $\frac{{\rm d}s_{\perp}}{{\rm d}X} = \frac{\rm A}{X_{\rm v}}$. The change of $s_{\perp}$ over an atmospheric depth of $\sim 100 \;$ g cm$^{-2}$ for the simulations (mixed) generated with the QGSJet model are $\frac{\rm A}{X_{\rm v}}\times 100 = \frac{0.31}{1000}\times 100 \sim 0.031$ and $\sim 0.033$ for two $N_{\rm e}$ ranges such as $(5.7\textendash 9.5) \times 10^{3}$ and $(2.0\textendash 5.0) \times 10^{4}$ respectively. At mountain altitude, the Mount Norikura group also reported similar results [49].  

In order to check the influence of the hadronic interaction models on the results obtained, we compare $s_{\perp}$ versus $\sec\Theta$ variations for two different models, QGSJet and SIBYLL, which are shown in the Figure 3. The SIBYLL yields comparatively a higher value for $s_{\perp}$ than QGSJet but the nature of dependence of the $s_{\perp}$ parameter on atmospheric depth is found similar in both the hadronic models. The discrepancy may come from the cross-section for proton-air collisions that rises more rapidly versus energy in SYBILL than in QGSJet.

\begin{figure}
\centering
\includegraphics[width=0.5\textwidth,clip]{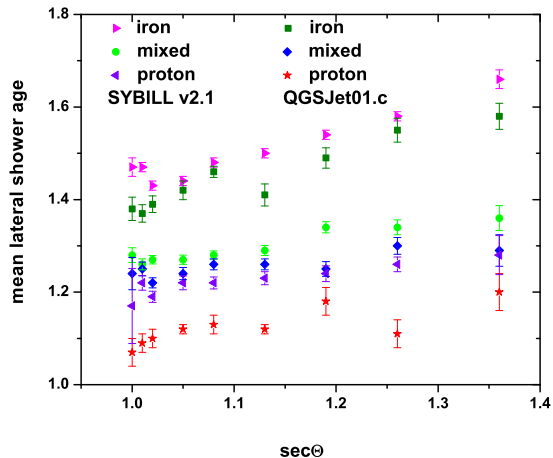} \hfill 
\caption{Variation of $s_{\perp}$ with $\sec\Theta$ for two different interaction models, QGSJet and SIBYLL corresponding to the $N_{\rm e}$ interval $(1.8\textendash 3.0) \times 10^{3}$ at NBU level.}
\end{figure}
\begin{figure}
\centering
\includegraphics[width=0.5\textwidth,clip]{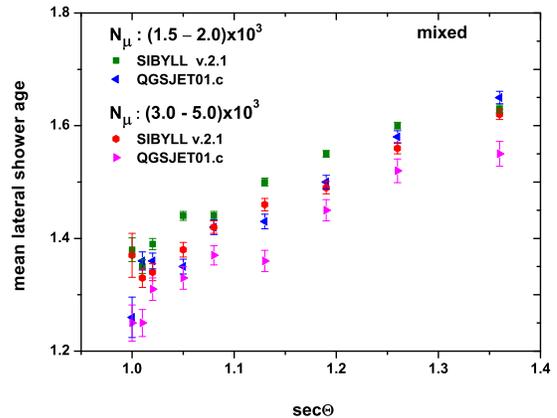} \hfill 
\caption{Variation of $s_{\perp}$ with $\sec\Theta$ for two muon size intervals using two high-energy hadronic interaction models at NBU level.}
\end{figure}

The variation of $s_{\perp}$ with $\sec\Theta$ is also studied for two different $N_{\mu}$ bins, which are shown in the Figure 4. We have included possible effects arising from the employment of different high-energy hadronic models into the results. As found in the Figure 3, the SYBILL still maintains a little higher value for $s_{\perp}$ relative to the QGSJet. 

\subsection{Correlation between $s_{\parallel}$ and $s_{\perp}$}
In Monte Carlo simulations the development of EM cascades towards the ground is calculated with the EGS4 code. The $s_{\parallel}$ parameter in terms of the numerical values obtained from the EGS4 simulation code attributes a  more refined procedure than what obtained from the EM cascade theory. Capdevielle et al. implemented an approximate formula for the $s_{\parallel}$ parameter for the EGS4 code [50-51]: 
\begin{equation}   
s_{\parallel} = \exp\left[0.67\times \left\{1+\frac{\alpha}{X}-\tau \right\}\right], 
\end{equation}
where $N_{\rm max}$ is the shower size corresponding to $X_{\rm max}$ of a shower and $\alpha = \ln\frac{N_{\rm max}}{N_{\rm e}}$, and $\tau = \frac{X_{\rm max}}{X}$.
The output data file is used to obtain the $s_{\parallel}$ parameter employing the equation (10). Such $s_{\parallel}$ parameter is used for comparison [50-51] with the predicted value by the cascade theory in approximation B.

\begin{figure}
\centering
\includegraphics[width=0.5\textwidth,clip]{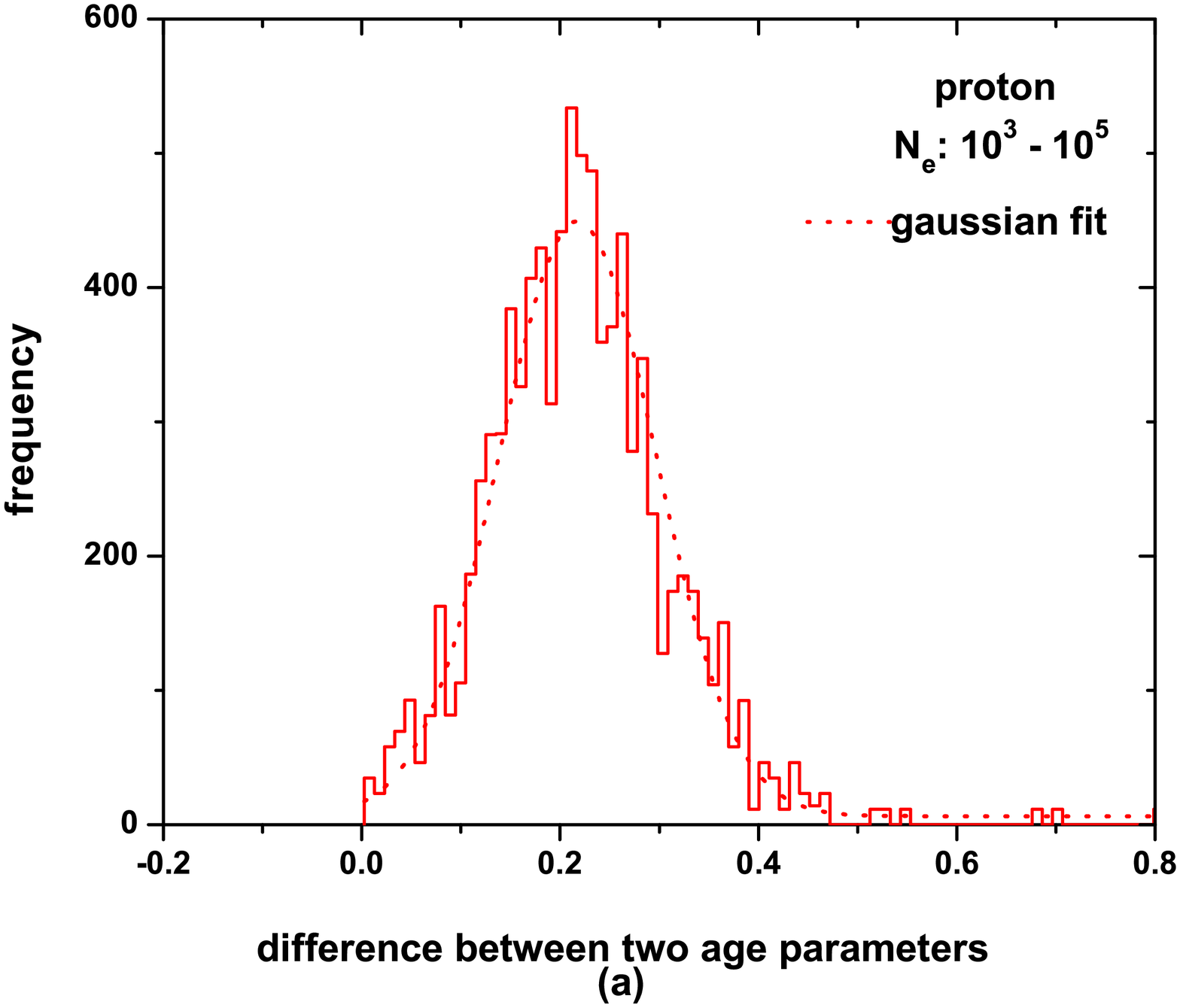} \hfill 
\includegraphics[width=0.5\textwidth,clip]{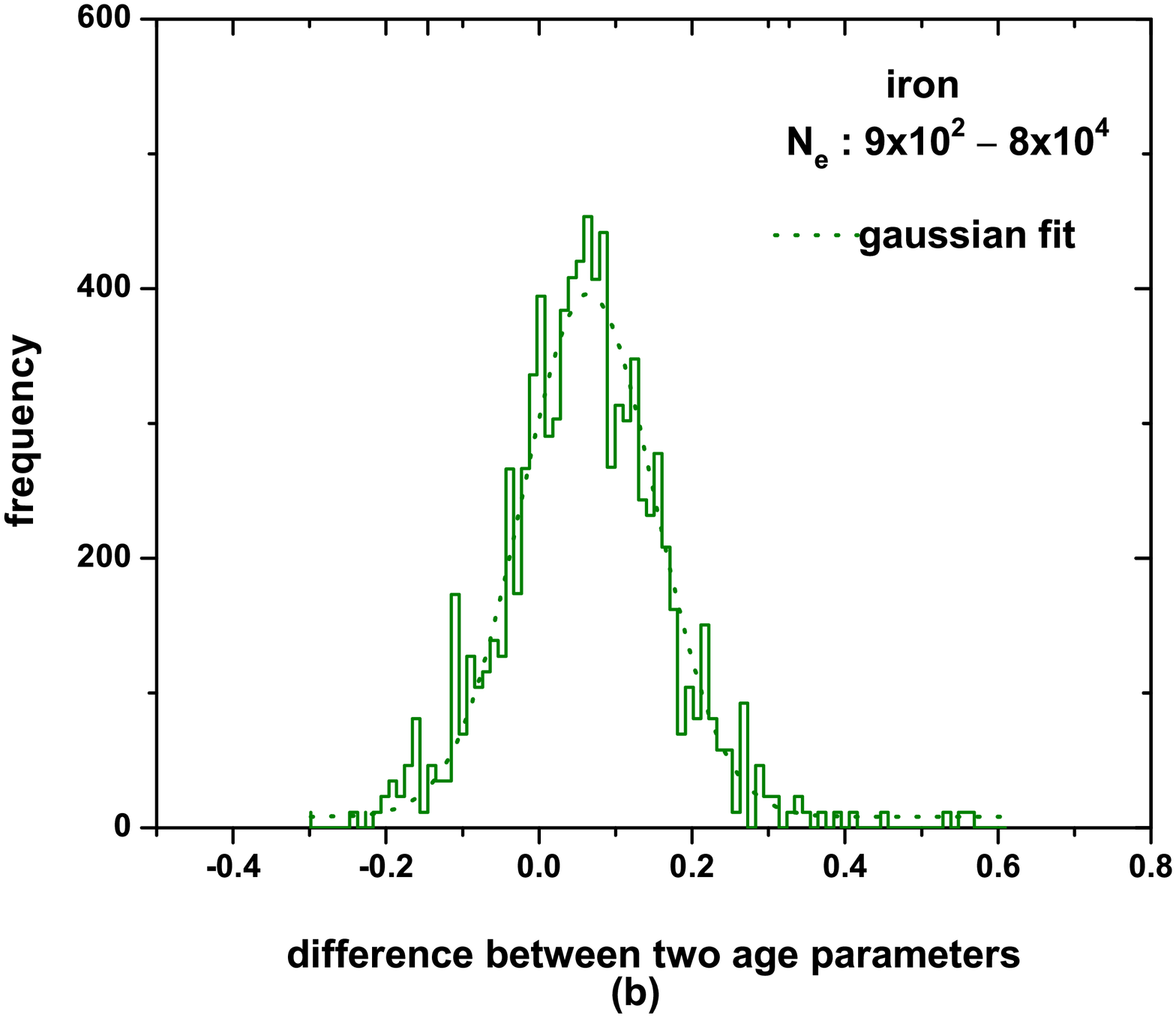} \hfill
\caption{Distributions of differences between $s_{\parallel}$ and $s_{\perp}$ at NBU level.}
\end{figure}

Hence, the shower age $s_{\parallel}$ can be estimated knowing both $N_{\rm max}$ and $X_{\rm max}$. The difference between the two age parameters, $s_{\parallel}$ and $s_{\perp}$, is studied. A frequency distribution of differences between two age parameters is given in the Figure 5a for proton induced showers. It is revealed from the figure that the frequency distribution peaks around $0.22\pm{.002}$ which is consistent with the early observations [31]. But, for iron showers the peak in the distribution occurs around a relatively lower value of $0.06\pm{.003}$, and is shown in the Figure 5b.    

\subsection{Distribution of $s_{\perp}$}
The distributions of $s_{\perp}$ are obtained from fitted showers for the  primary energy range $3\times 10^{14}\textendash 3\times 10^{16}$ eV and are  shown in Figure 6a and 6b. A Gaussian distribution function does not nicely fit these distributions. Deviations from the Gaussian nature become visible or distributions become asymmetric when the observational level is much deeper than the depth of the shower maximum, which is the case for both KASCADE and NBU levels. In those cases, fitting quality can be improved by an Extreme Value Distribution (E.V.D.) function. The E.V.D. probability density function for the $s_{\perp}$ parameter has the following form,
\begin{equation}   
f_{\rm EVD}(s_{\perp}) = \frac{1}{\sigma_{s_{\perp}}}\;\exp\left[\pm\frac{(\mu - s_{\perp})}{\sigma_{s_{\perp}}}-\exp\left\{\pm\frac{(\mu - s_{\perp})}{\sigma_{s_{\perp}}}\right\}\right].
\end{equation} 
The parameters $\mu$ and $\sigma_{s_{\perp}}$ are related to the average size $\overline{s_{\perp}}$ and its variance ${\rm V}_{s_{\perp}}$ by $\overline{s_{\perp}}=\mu \pm 0.577\;\sigma_{s_{\perp}}$ and ${\rm V}_{s_{\perp}} = 1.645{\sigma_{s_{\perp}}}^2$. 

In Figure 6a and 6b, we have shown histograms for $s_{\perp}$ obtained from our Monte Carlo simulations. In these figures, the frequencies reflect the intensities of primaries around the knee. The long tail in the left wing of Figure 6a for example corresponds to very penetrating proton showers of low primary energy and conversely the steep right wing contains showers interacting at very high altitude whereas the central part is populated by showers with an individual maximum close from average maximum at each energy. Similar features concerning the case of pure iron component (Figure 6b) with a general reduction in $\sigma_{s_{\perp}}$ value of the fluctuation. As an example, the analysis of the Figure 6b by the E.V.D. yields $\sigma_{s_{\perp}} = 0.07$ and $\mu = 1.495$. From the Figure 6a, we will obviously obtain different set of values for these parameters. Hence, it reveals that these frequency distributions fitted by the E.V.D., could be useful for extracting information on EAS initiated primaries around the knee.

\begin{figure}
\centering
\includegraphics[width=0.5\textwidth,clip]{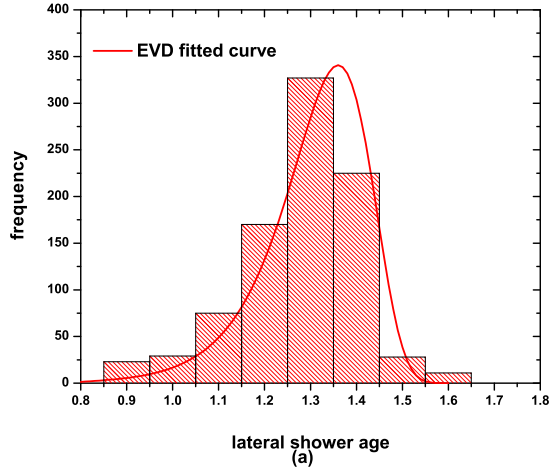}  
\includegraphics[width=0.5\textwidth,clip]{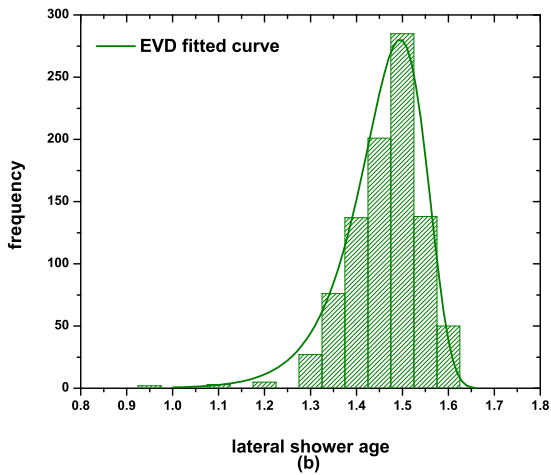} 
\caption{Distributions of $s_{\perp}$ for vertically incident showers at the NBU level for the energy range $E = 3\times 10^{14}\textendash 3\times 10^{16}$ with the QGSJet model; (a) proton and (b) iron. Fits are made by the EVD function. For (a) $\mu \approx 1.38$ and $\sigma_{s_{\perp}} \approx 0.1$ and (b) $\mu \approx 1.495$ and $\sigma_{s_{\perp}} \approx 0.07$.}
\end{figure}
\begin{figure}
\centering
\includegraphics[width=0.5\textwidth,clip]{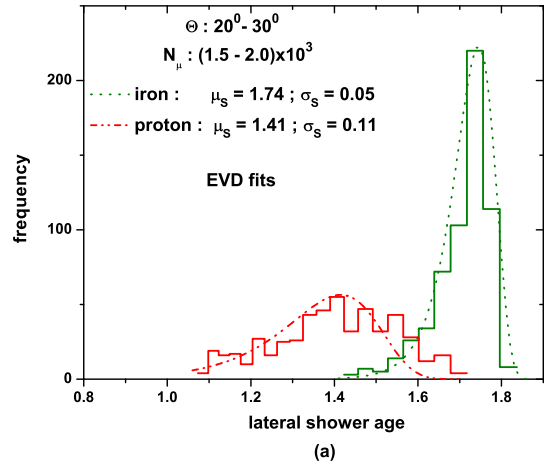} \hfill 
\includegraphics[width=0.5\textwidth,clip]{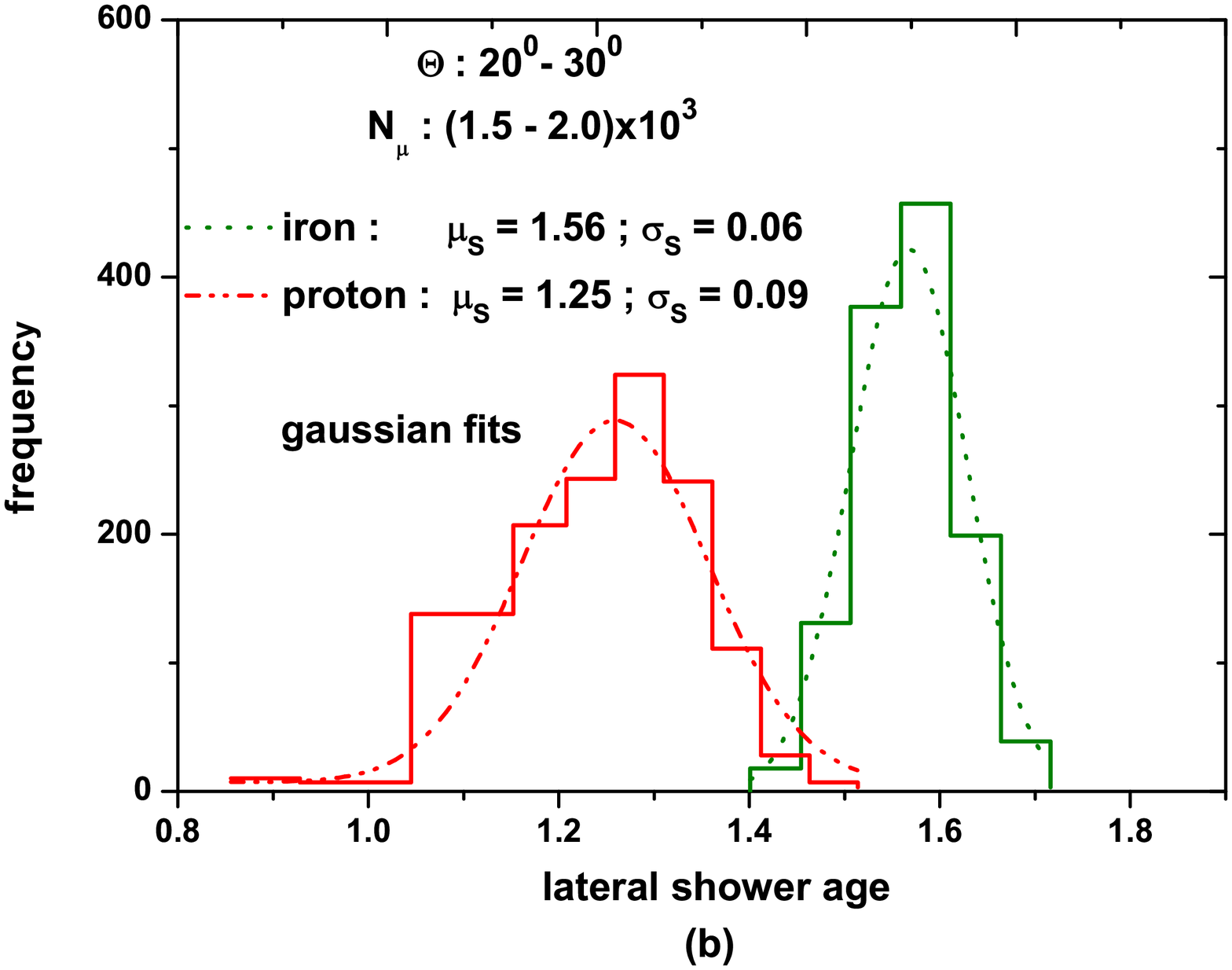} \hfill
\caption{In (a) the distributions of $s_{\perp}$ are shown for proton and iron showers at the NBU level (sea level) with EVD fits. In (b) the distributions are shown at ARGO-YBJ level (mountain level) with Gaussian fits. The QGSJet model is used entirely.}
\end{figure}
\begin{figure}
\centering 
\includegraphics[width=0.5\textwidth,clip]{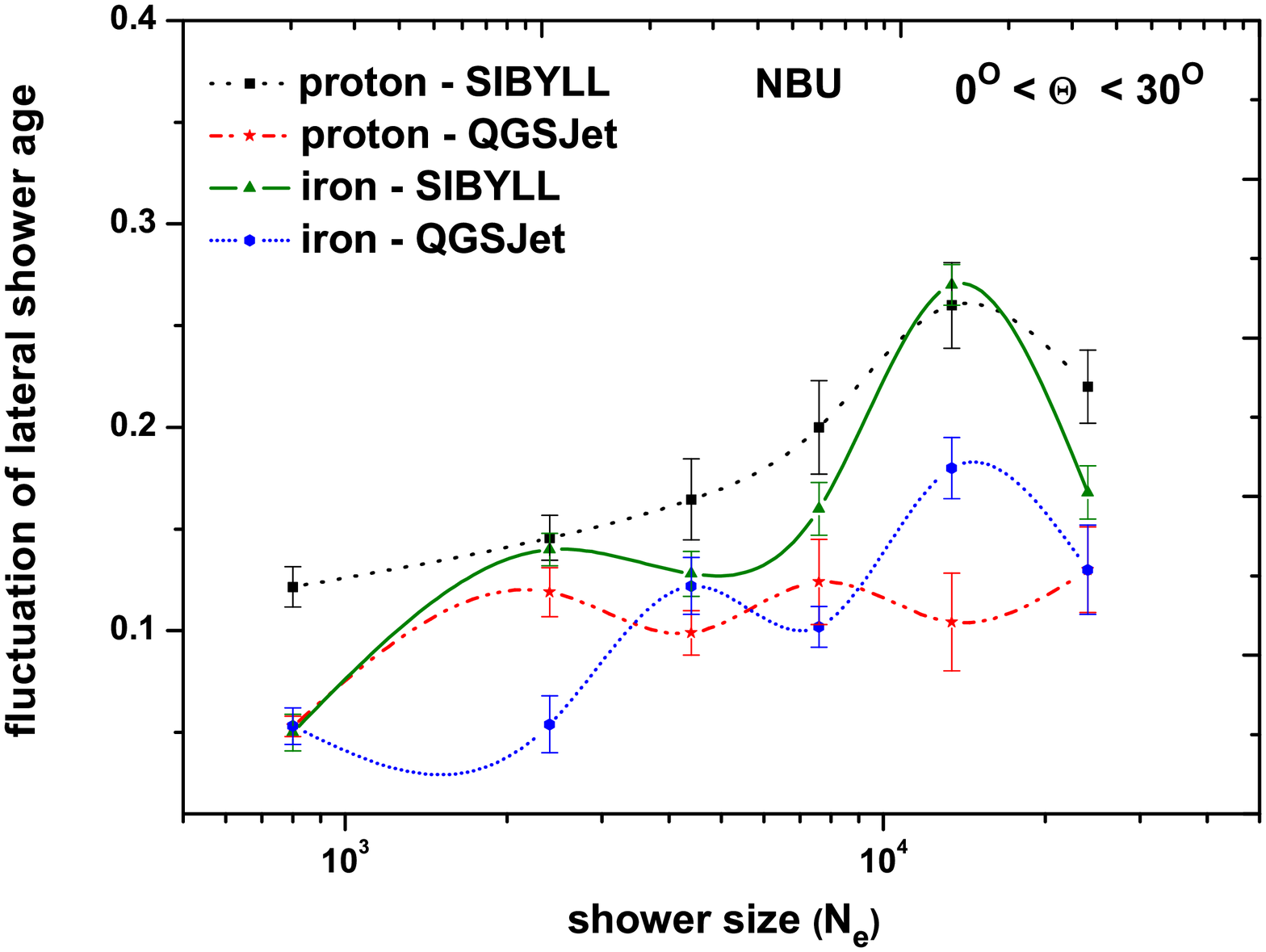} \hfill
\caption{Fluctuations in $s_{\perp}$ and their variation with $N_{\rm e}$ at the NBU level.The lines are only a guide for the eye.}
\end{figure}

In the Figure 7a, a small $N_{\mu}$ range is considered in order to distinguish between frequency distributions of $s_{\perp}$ for proton and iron induced showers at the NBU level. This study could be very effective to extract information on the mass composition of PCRs in data at sea levels. We describe the typical histograms here by the E.V.D. function also. 

Frequency distributions of $s_{\perp}$ for proton and iron showers in the same $N_{\mu}$ range like Figure 7a are presented through the Figure 7b at a high altitude location in order to correlate the $s_{\perp}$ with longitudinal development of an EAS (here, we have used our analyzed simulated data from the work in [52]). As the mountain level is not much deeper than the depth of the shower maximum, the asymmetries of the $s_{\perp}$ distributions will be marginal and can be conveniently described by the Gaussian distribution function. Therefore, the $s_{\perp}$ histograms in the Figure 7b are fitted by the Gaussian distribution function. Distributions in the Figure 7b suggest that the smaller $s_{\perp}$ values account for the smaller extents of EAS longitudinal developments and showers are regarded as young. Comparison of mean values either for proton or iron initiated showers between the Figures 7a and 7b speak for themselves. This is in agreement with the prediction from the EM cascade theory. The EM theory argues that the LDF shape parameter will not differ much from its value at $X_{\rm {max}}$ for altitudes relatively nearer to $X_{\rm {max}}$ and the condition $s_{\perp} \approx s_{\parallel} \approx 1$ is also valid. 

The fluctuations in $s_{\perp}$ are larger for proton initiated showers in compared to those initiated by heavier primary as revealed from all the figures in Figure 6 and 7.

\begin{figure}
\centering 
\includegraphics[width=0.5\textwidth,clip]{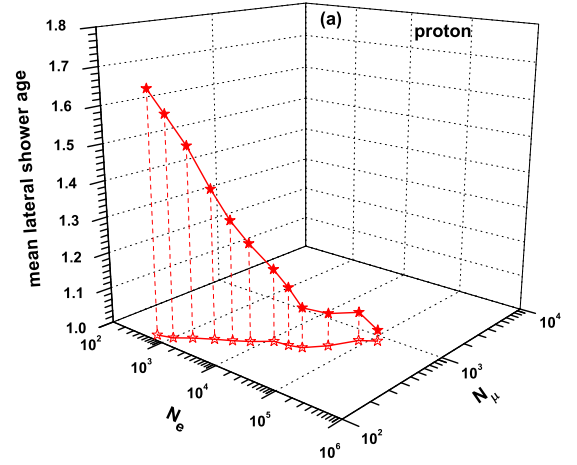}\hfill 
\includegraphics[width=0.5\textwidth,clip]{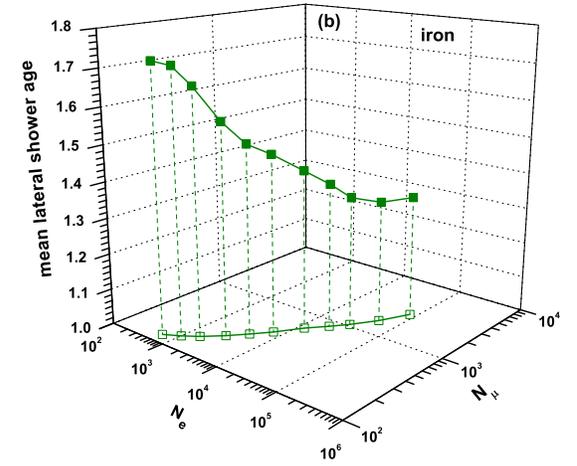}\hfill
\includegraphics[width=0.5\textwidth,clip]{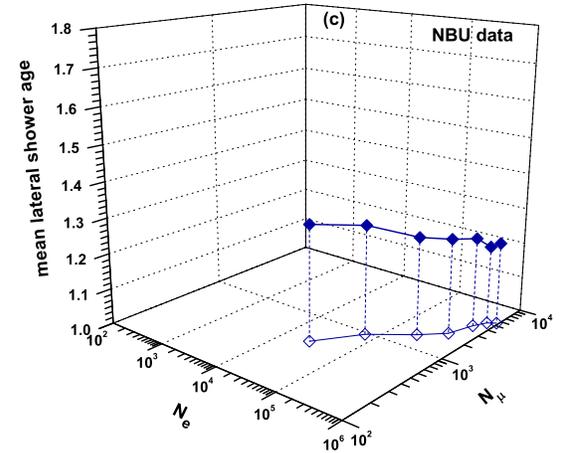}\hfill
\caption{3-Dimensional plot using shower size, muon size and lateral shower age parameters at NBU level: (a) Pure proton and (b) Pure iron showers, generated using the QGSJet model. NBU data are given in (c). Muons are considered from $4\textendash35$ m core distance with $E_{\mu}^{\rm Th.} \geq 2.5$ GeV. Projection on the ${\rm X}\textendash {\rm Y}$ plane showing the corresponding $N_{\rm e}\textendash N_{\mu}$ curve.}
\end{figure} 

\subsection{The fluctuation of $s_{\perp}$ as a function of shower size}
The fluctuations of $s_{\perp}$ in different smaller shower age bins are estimated and plotted with increasing mean $N_{\rm e}$ for proton and iron showers in the Figure 8. It is found that fluctuations in $s_{\perp}$ are larger for EASs induced by proton than iron, except at lower energies. This is in accordance with the predictions made in [53]. However, the SIBYLL introduces little higher fluctuations to $s_{\perp}$ in compare to fluctuations contributed by the QGSJet model for any particular type of primary. It appears from the Figure 8 that fluctuation (i.e. the R.M.S of variance of $s_{\perp}$) has essentially zero mass-separation power but rather a quite large model-dependence.

\begin{figure}
\centering 
\includegraphics[width=0.5\textwidth,clip]{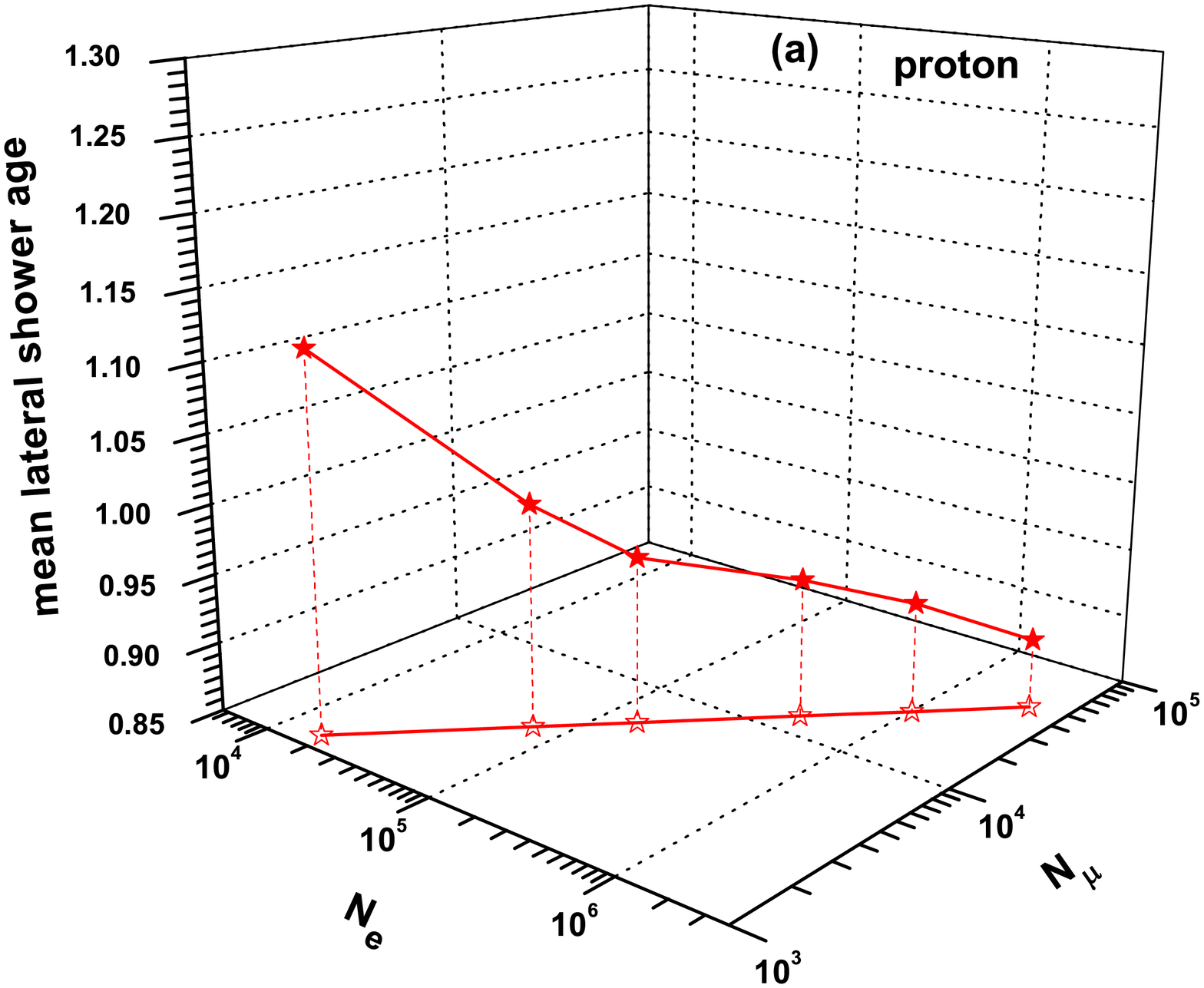}\hfill 
\includegraphics[width=0.5\textwidth,clip]{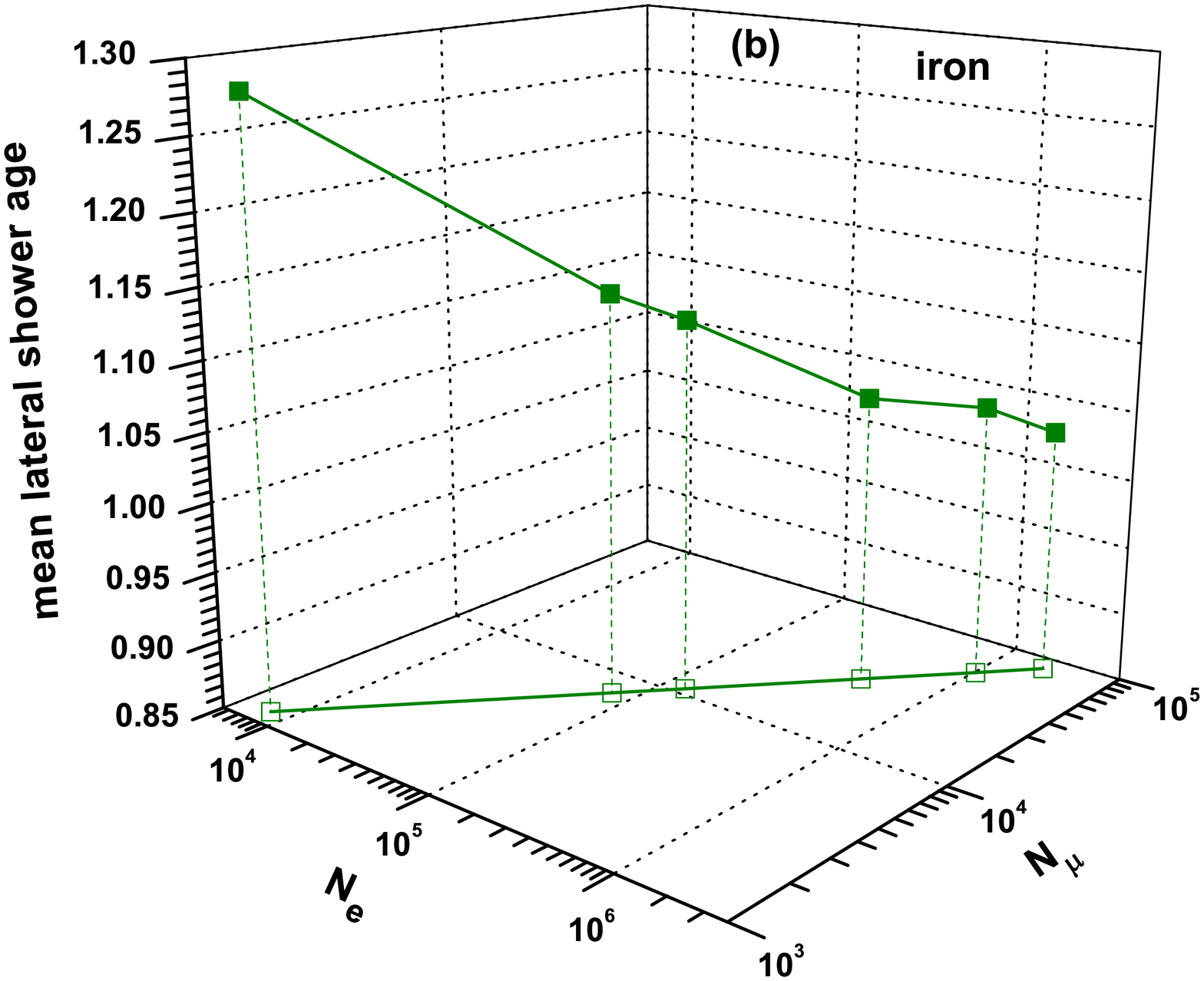}\hfill
\includegraphics[width=0.5\textwidth,clip]{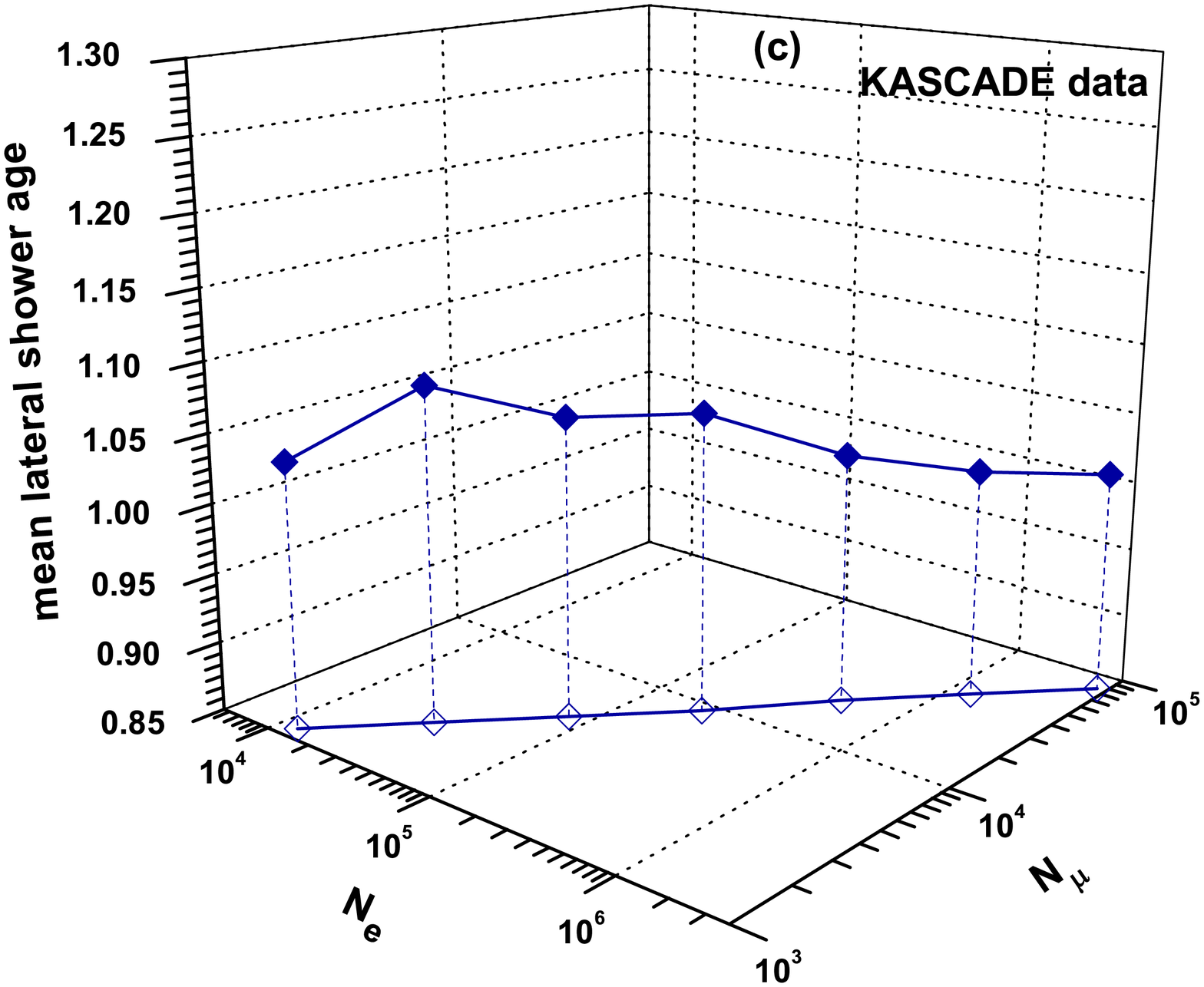}\hfill
\caption{3-Dimensional plot using shower size, muon size and lateral shower age parameters at KASCADE level: (a) Pure proton and (b) Pure iron showers, generated using the QGSJet model. KASCADE data are given in (c). Projection on the ${\rm X}\textendash {\rm Y}$ plane showing $N_{\rm e}\textendash N_{\mu}$ curve.}
\end{figure}

\subsection{Variation of $s_{\perp}$ simultaneously with $N_{\rm e}$ and $N_{\mu}$: A multi-parameter approach} 

Muon content of EAS is generally used to extract information on the composition of CR primaries. However, due to uncertainties in the interaction model and the difficulties associated with the solutions to the EAS inverse problem, separating the primary energy spectra of elemental groups remains ambiguous. Proton and iron initiated showers may be better separated employing $s_{\perp}$ and $N_{\mu}$ simultaneously through 3-dimensional plot of $N_{\rm e}$ versus $N_{\mu}$ and $s_{\perp}$ than in either of its 1-dimensional projections. It is found that accuracy of determining the nature of primary species increases with the simultaneous use of $s_{\perp}$ and $N_{\mu}$ against $N_{\rm e}$ or $E$.

In Figures 9a, 9b, 10a and 10b, we plot the 3-dimensional curve between $\overline{s_{\perp}}$, $\overline{N_{\rm e}}$ and $\overline{N_{\mu}}$obtained from the simulations for both proton and iron primaries at NBU and KASCADE levels. The corresponding observational results of the two experiments are also shown through the Figures 9c and 10c. 

For the NBU data, measurements of various observables are made over a limited energy window [38]. The detecting efficiency of the NBU array within its detecting area is greater than $90\%$ for showers of size range $\sim{10^4}$ to $10^6$. The array achieves $\sim 1.1^{\rm o}$ angular resolution in the $\Theta$ range $0^{\rm o} - 40^{\rm o}$. However at higher zenith angles ($> 40^{\rm o}$, the resolution becomes increasingly poorer, it becomes $4^{\rm o}$ around $55^{\rm o}$ zenith. The energy resolution corresponds to shower size resolution in the NBU experiment. The accuracy of the size estimation depends on fluctuations in the cascade development in the atmosphere. Fluctuations of size are studied employing Monte Carlo simulation. Fluctuation of size estimation increases with the decrease of angular resolution i.e. with increasing zenith. 

The muon detecting system could detect and measure muons associated with an EAS up to a momentum 500 GeV/c. The lower cutoff energy of muons, provided by the lead and concrete absorbers placed above the pair of magnet spectrograph is 2.5 GeV. The measurement for the $N_{\mu}$ are made considering muon density data within a limited radial distance range of 4$\textendash$35 m only. To get better comparison between simulation and experiment in NBU, we have sorted out muons from the region of $r = 4\textendash 35$ m range with momenta $p_{\mu} \geq 2.5$ GeV/c from generated events. A comparison among three figures in 9a, 9b and 9c indicates that the NBU data though obtained in a small energy window but still hint for a mixed composition across the knee. The result shows that the 3-dimensional plot is more distinct although we have constraints on the NBU muon data for the PCR composition study than its 2-dimensional projection through the $N_{\rm e} \textendash N_{\mu}$ curve.

In the case of KASCADE data, we have estimated $s_{\perp}$ from their measured lateral distribution [48,54-55] employing our fit procedure for a given $N_{\rm e}$ bin. The corresponding $N_{\mu}$ are extracted from the $N_{e}\textendash N_{\mu}$ curve [56]. The comparison of simulated results with data from the KASCADE experiment on the basis of the 3-dimensional plots (Figures 10a, 10b, 10c) indicate for a noticeable change in primary composition towards heavier domination as $N_{\rm e}$ increases across the knee. The KASCADE group also reached the similar conclusion using the slope parameter in the modified NKG LDF instead of $s_{\perp}$ [48]. The 3-dimensional plot over the 2-dimensional $N_{\rm e} \textendash N_{\mu}$ curve seems advantageous for the PCR mass composition study so far the KASCADE results are concerned. 

In the KASCADE experimental data, the error in $s_{\perp}$ estimation remains within $0.06$, whereas for the NBU data the error was found to be within $0.04$.

\section{Conclusions}

The present work, with particular focus on the $s_{\perp}$ parameter, has yielded the following conclusions.

(1) The $s_{\perp}$ parameter correlates with the stage of shower development on a statistical basis; average of this parameter increases as air showers traverse an increased thickness of atmosphere. This expected behavior is found for two different hadronic interaction models, QGSJet and SIBYLL. The NBU data agree with the simulation and also some early works reported nearly 4 decades ago.

The frequency distributions of differences between $s_{\perp}$ and $s_{\parallel}$ also establish the fact that longitudinal and lateral developments of EAS are strongly correlated through their shape parameters. Such distributions and their central values are found mass sensitive.

(2) The $s_{\perp}$ parameter takes higher values for iron initiated showers as compared to proton showers. This means that lateral distribution of electrons for iron initiated showers is flatter (or older) relative to that of proton showers (younger). The slope of $s_{\perp}$ versus atmospheric depth curve is, however, more or less the same for proton and iron showers. We have also noticed that for mixed primaries the slope of $s_{\perp}$ versus atmospheric depth curve is found smaller compared to pure primary initiated EAS. The observation by the NBU air shower array on this aspect [48] indicates for a dominant lighter (proton) species up to at least around the knee energy region.

(3) The $s_{\perp}$ parameter has been found to decrease with the increase of both $N_{\rm e}$ and $N_{\mu}$ at least for simulated pure proton. But for simulated iron showers the rate of fall is quite slow, becomes uniform at last in the NBU simulation. However, the variation for experimental data shows a gradual rise instead and, in particular in the KASCADE data (from proton proximity to iron). This clearly indicates a domination of relatively heavy primaries beyond the knee. Due to certain limitations of the NBU experiment, the data do not exhibit such distinct signature but favour a mixed composition across the knee.\\

Many EAS experiments measured the CR mass composition using the conventional approach of implementing the variation of $N_{\mu}$ against $N_{\rm e}$, which we have obtained through the 2-dimensional projection on the ${\rm X}\textendash {\rm Y}$ plane in each case. Because of a generic feature of EAS development, the total muon number on the ground varies from one species to other and that makes the $N_{\mu}$ as a possible mass sensitive parameter. But, due to limited area covered by muon detectors in an EAS experiment, a truncated size of muons ($N_{\mu}^{\rm {tr.}}$) is obtained. On the other hand, there is no air shower model that can accurately describe $N_{\mu}$ right now (e.g. EPOS gives a little higher $N_{\mu}$ compared to QGSJet and others). Therefore, a precise mass composition study based on $N_{\mu}$ only is still not possible. A comparison among three projected 2-dimensional plots for either the KASCADE or NBU is in accordance with the above prediction involving muons.

(4) The frequency distribution and fluctuation of the $s_{\perp}$ parameter might be useful to predict the nature of the EAS primary. However, the variation of $\sigma_{\perp}$ with $N_{\rm e}$ does not look effective for the measurement of CR mass composition. 

\section*{Acknowledgment} The authors would like to thank the anonymous reviewer for detailed and useful comments in improving the quality of the manuscript. RKD acknowledges the financial support from the SERB, Department of Science and Technology (Govt. of India) under the Grant no. EMR/2015/001390.  

\appendix*
\section{Elements on the theory of the EM cascade}

The one dimensional (1D) and the three dimensional (3D) theories are distinguished based on whether the theory addresses just longitudinal or both lateral and longitudinal shower development. In the analytical approach, the time distribution can be derived from the solution of the 3D model providing the densities of the particles and their energy distributions. The four dimensional (4D) simulation is more generally reserved to the Monte Carlo approach where electrons and photons are followed simultaneously in space-time coordinates [40].  

\subsection{Approximations in the theoretical model of the cascade diffusion equations}

The approximation A  neglects the ionization losses taking into account the radiation process and the pair creation process. It allows to establish from the elementary gains and losses in particles and energy. In the case of the 1D theory, the most simple system of transport equations which are perfectly symmetric [15], is given by

\begin{equation}  
\frac{\delta \boldsymbol{\pi}}{\delta X} =  -{\rm A'}\boldsymbol{\pi} + {\rm B'} \boldsymbol{\gamma}\; 
\end{equation}

\begin{equation}
\frac{\delta \boldsymbol{\gamma}}{\delta X} =  -\sigma_{0} \boldsymbol{\gamma} + {\rm C'}\boldsymbol{\pi}\;
\end{equation}

We use here the notations of Nishimura in [15], $\boldsymbol{\pi}(E,X)dE$ and $\boldsymbol{\gamma}(E,X)dE$ representing respectively the average numbers of electrons and photons with energy between $E$ and $E+dE$ at a depth $X$ in radiation lengths. The integral operators $\rm{A'}$ and $\rm{B'}$ correspond respectively to the losses in electron number by bremsstrahlung and gains by pair productions, whereas $\sigma_{0}$ represents the cross section for pair production (but, $-\sigma_{0}\gamma$ accounts the loss of photons through pair production). The integral operator $\rm{C'}$ corresponds the gain in number of photons from electron bremsstrahlung. 

The approximation A is better adapted to the growing phase of the cascade and to high energy photons and electrons. The approximation B is more realistic one. In addition to the approximations stated under approximation A, it also incorporates the effect of the ionization loss $\epsilon \frac{\delta \boldsymbol{\pi}}{\delta E}$ , thus replacing the first term of the system (A1) by 

\begin{equation}
 \frac{\delta \boldsymbol{\pi}}{\delta X} =  -{\rm A'}\boldsymbol{\pi} + {\rm B'}\boldsymbol{\gamma} + \epsilon\frac{\delta \boldsymbol{\pi}}{\delta E}\;
\end{equation} 

In parallel, the 3D diffusion equations can be inferred after introducing the functions $\boldsymbol{\pi}(E,\boldsymbol{r},\boldsymbol{\theta})$ and $\boldsymbol{\gamma}(E,r,\boldsymbol{\theta})$. For instance, in the case of the approximation B they take the simple form in the so called Landau approximation:

\begin{equation}
\frac{\delta \boldsymbol{\pi}}{\delta E} + \boldsymbol{\theta} \frac{\delta \boldsymbol{\pi}}{\delta \boldsymbol{r}} =  -{\rm A'}\boldsymbol{\pi} + {\rm B'}\boldsymbol{\gamma} + \epsilon \frac{\delta \boldsymbol{\pi}}{\delta E} +  \frac{E_{s}^2}{4E^2} (\frac{\delta^2 \boldsymbol{\pi}}{\delta  \theta_{1}^2} + \frac{\delta^2 \boldsymbol{\pi}}{\delta \theta_{2}^2}) \; 
\end{equation}

\begin{equation}
\frac{\delta \boldsymbol{\gamma}}{\delta X} + \boldsymbol{\theta} \frac{\delta \boldsymbol{\gamma}}{\delta \boldsymbol{r}} =  -\sigma_0\boldsymbol{\gamma} + {\rm C'}\boldsymbol{\pi}
\end{equation}

The more general differential description in Approximation B without Landau simplification is obtained by replacing in this last system equation (A5) by the expression

\begin{equation}
\frac{\delta \boldsymbol{\pi}}{\delta E} + \boldsymbol{\theta} \frac{\delta \boldsymbol{\pi}}{\delta \boldsymbol{r}}  =  -{\rm A'}\boldsymbol{\pi} + {\rm B'} \gamma + \epsilon \frac{\delta \boldsymbol{\pi}}{\delta E} +  \int [\boldsymbol{\pi}(\boldsymbol{\theta} - \boldsymbol{\theta'})- \boldsymbol{\pi}(\boldsymbol{\theta'})] \sigma(\boldsymbol{\theta'}) {\rm d}\boldsymbol{\theta'} 
\end{equation}  

The multiple coulomb scattering governs the lateral deflections submitted by the electrons passing through the atmosphere. The small angle approximation owing a simple expression of the mean scattering angle when the electron passes through an elementary thickness $\rm{d}X$ is an important step to express the equilibrium described by the equations (A4-A5) and (A6) , especially (A4-A5) with the Landau approximation.

\subsection{Approximations in the numerical treatment}

The differential equations (A4) and (A5) were solved after application of the Hankel transform to {r}, Mellin transform to $E$ and Laplace transform to $X$ [11,15]. Re-transformations from the solutions expressed in terms of complex functions required further approximations, the most important being the saddle point method. The final results expressed in numerical densities were again fitted with a tolerable agreement with the so called NKG formula (2) following an earliest Nishimura formula.
 
Another approach to solve the system of equations (A4-A5) was performed by the method of adjoint equations [12-13] and the resulting structure functions were found steeper. Here also several numerical approximations had to make to get the solution. 

Surprisingly, $\rho_{\rm {EM}}$(r) which have not been checked above $10^{17}$ eV was used \textit{in extenso} to calculate the radio synchrotron emission in giant extensive air showers [57]: recalculating those densities with CORSIKA-EGS4, we observed that the discrepancies remain small for axis distances lower than 3$r_{\rm m}$ containing fortunately the largest part of the source of radio emission or fluorescence in EAS.  
 
\subsection{Shower age parameter in longitudinal and lateral developments}

The age parameter $s$ was first derived from the solution of equations (A1-A2) using the Mellin transformation, $s$ being the variable in the complex plane.  The total number of electrons (obtained by the inverse Mellin transformation) is obtained for $s = \overline s$ following the relation
\begin{equation}
\lambda_{1} '(\overline s)X + \log \frac{E_0}{E}  - \frac{1}{\overline s} = 0.
\end{equation}

Taking into account the elementary solutions (where $\lambda_1 '(\overline s$) is a function varying slowly), the relation between $s$ and $X$ was derived from the approximation at cascade maximum 
\begin{equation}
\frac{\delta \overline s}{\delta X}(\lambda_1 '(\overline s)X + \log \frac{E_0}{E}  - \frac{1}{\overline s}) +\lambda_1 (\overline s) = 0
\end{equation} 
and the general properties of $s$ were established by the fact that the maximum of the cascade is at $s = 1$, the cascade is developing (growth) when $s \le 1$, whereas it is decaying (absorption) if $s \ge 1$.

Those considerations in approximation B are brought through the relation (8) for $s_{\parallel}$, the so called Greisen formula for longitudinal development, along with the relation (5). One of the most clear presentation of the qualitative relation between $s_{\parallel}$ and the lateral profile of the cascade under approximation B was demonstrated by Cocconi [14] showing that a lateral distribution is becoming flatter when $s \ge 1$, of course for $s_{\parallel} = s_{\perp}$. The case of relation (4) for $s_{\parallel}$ was also considered by Cocconi and Nishimura to take into account some effects of density resulting of the atmospheric inhomogeneity. Relation (5) has several advantages passing from an asymptotic tendency near axis when $r \rightarrow 0$ following $r^{s-2}$ to a steeper power law when $r \rightarrow \infty$. The NKG function is based on the Eulerian Beta function, $\rm {B}(u,v)$, taken here in the case of cylindrical symmetry, from~:
\begin{eqnarray}
 N_{\rm e} = \int_{0}^{\infty} 2 \pi r \rho_{\rm{NKG}}(r) {\rm d}r \\
 = 2 \pi {\rm C}(s_{\perp}) \int_{0}^{\infty} ({r\over r_{\rm m}})^{s_{\perp}-1} ({r\over r_{\rm m}} +1 )^{s_{\perp}-4.5} {\rm d}({r\over r_{\rm m}}),
\end{eqnarray}

where appears the classical form~:
\begin{equation}
\rm{B}(u,v) = \int_{0}^{\infty} {{y^{u+1} \over {(1+y)^{u+v}}}} {\rm d}y,
\end{equation}
for $y= {r \over r_{0}}$, u = $s_{\perp}-2$, v = $6.5-2s_{\perp}$.

This normalization via (A9) gives the opportunity to link one single density to $N_{\rm e}$ as 
\begin{equation}
\rho_{\rm{NKG}}(r) = {N_{\rm e} \over r_{\rm m}^2 } {\rm f}_{\rm{NKG}}(r)
\end{equation}
where $s_{\perp}$ must be a constant (versus $r$) corresponding to a fixed value of $X$ in relation (5), otherwise expressing the integrations in term of Euler Beta function is not valid in general.\\

\end{document}